\documentclass[floatfix,lengthcheck,showpacs,amssymb,amsmath,amsfonts,twocolumn,nofootinbib,longbibliography,superscriptaddress]{revtex4-1}
\usepackage{amsfonts,amsmath,wasysym,epsfig,graphicx,verbatim,color,subfigure,graphicx}
\usepackage{amsmath}
\usepackage{latexsym}
\usepackage{amssymb}
\usepackage{amsfonts}
\usepackage{mathtools}
\usepackage{bm}
\usepackage{color}
\usepackage{float}
\usepackage{tikz}
\usepackage{adjustbox}
\usepackage{array}
\usepackage{soul}
\usepackage{appendix}
\usepackage{physics}
\usepackage{braket}
\usepackage{xcolor}
\usepackage{booktabs}
\usepackage{siunitx}
\usepackage[margin=1in]{geometry}
\usepackage[none]{hyphenat}
\usepackage{orcidlink} 
\usepackage{lineno}
\usepackage{gensymb}
\usepackage{siunitx}

\usepackage{booktabs} 

\def\be{\begin{equation}}
\def\ee{\end{equation}}
\def\bea{\begin{eqnarray}}
\def\eea{\end{eqnarray}}

\def \refFreq {20}

\begin{document}

\title{Potential science with GW250114 -- the loudest binary black hole merger detected to date}

\author{{Aleyna Akyüz}\orcidlink{0000-0001-7248-951X}}
\affiliation{Department of Physics, Syracuse University, Syracuse, NY 13244, USA}

\author{{Alex Correia}\orcidlink{0009-0000-9756-9794}}
\affiliation{Department of Physics, Syracuse University, Syracuse, NY 13244, USA}

\author{{Jada Garofalo}\orcidlink{0009-0001-5699-2294}}
\affiliation{Department of Physics, Syracuse University, Syracuse, NY 13244, USA}

\author{{Keisi Kacanja}\orcidlink{0009-0004-9167-7769}}
\affiliation{Department of Physics, Syracuse University, Syracuse, NY 13244, USA}

\author{{Labani Roy}\orcidlink{0009-0004-8445-6212}}
\affiliation{Department of Physics, Syracuse University, Syracuse, NY 13244, USA}

\author{{Kanchan Soni}\orcidlink{0000-0001-8051-7883}}
\affiliation{Department of Physics, Syracuse University, Syracuse, NY 13244, USA}

\author{{Hung Tan}\orcidlink{0000-0001-9101-048X}}
\affiliation{Department of Physics, Syracuse University, Syracuse, NY 13244, USA}

\author{{Vikas Jadhav Y}\orcidlink{0009-0008-5970-9694}}
\affiliation{Department of Physics, Syracuse University, Syracuse, NY 13244, USA}

\author{{Alexander H. Nitz}\orcidlink{0000-0002-1850-4587}}
\affiliation{Department of Physics, Syracuse University, Syracuse, NY 13244, USA}

\author{{Collin D. Capano}\orcidlink{0000-0002-0355-5998}}
\email{cdcapano@syr.edu}
\affiliation{Department of Physics, Syracuse University, Syracuse, NY 13244, USA}
\affiliation{Physics Department, University of Massachusetts Dartmouth, North Dartmouth, MA 02747, USA}

\date{\today}
 
\begin{abstract}
On January 14, 2025 the LIGO interferometers detected a gravitational wave from the merger of two black holes, GW250114. Using publicly available information, we estimate that the signal-to-noise ratio (SNR) of GW250114 was $\sim 80$. This would make it three to four times louder than any other gravitational wave detected to date. GW250114 therefore offers a unique opportunity to make precise measurements of its source parameters and to test general relativity. In anticipation of its public data release, we analyze a set of simulated signals that have parameters similar to what we estimate for GW250114 and explore what new insights may be gained from this significant event. We investigate how well the component spins may be constrained, whether any eccentricity may be measured, what quasi-normal modes (QNMs) may be detected in the post-merger signal, how well the black hole area theorem may be constrained, and what constraints may be expected on sub-dominant inspiral-merger-ringdown modes. We find that it should be possible to measure a non-zero eccentricity at $20\,$Hz ($e_{20}$) if GW250114 has $e_{20} \gtrsim 0.05$. We also find that at least one overtone of the dominant QNM should be detectable in the ringdown of GW250114, with a Bayes factor of $O(10^3)$ after marginalizing over all timing uncertainties.

\end{abstract}

\maketitle

\section{Introduction}

The fourth observing run (O4) of Advanced LIGO~\cite{LIGOScientific:2014pky}, Virgo~\cite{VIRGO:2014yos}, and KAGRA~\cite{KAGRA:2018plz} began on May 24, 2023. Since then,\footnote{As of May 2025.} the network of gravitational-wave (GW) detectors has collected over a year of data~\cite{DetectorStatus,Akyuz:2025ype}. The detector's sensitivity during this period has been unprecedented: the luminosity distance at which a compact binary merger can be detected has more than doubled compared to the first observing run~\cite{DetectorStatus,LIGOScientific:2018mvr}. This has led to a corresponding increase in the number of detections; a binary merger is now detected every $\sim 3$ days, yielding over 200 GW observations to date~\cite{GraceDB,Akyuz:2025ype}.

Although the full O4 dataset has not yet been made public, the LIGO-Virgo-KAGRA (LVK) collaboration releases low-latency alerts for candidate events in near real-time~\cite{publicalert}. These events facilitate rapid electromagnetic follow-ups and provide preliminary information about source classification and source properties such as sky location. In a companion paper~\cite{Akyuz:2025ype}, we show that it is also possible to estimate the signal-to-noise ratio (SNR) and the chirp mass of these events using the publicly available information.

On January 14, 2025, the LVK Collaboration identified a compact binary merger event GW250114, likely originating from the coalescence of a binary black hole (BBH)\cite{GraceDB}. Based on our analysis of publicly available information~\cite{Akyuz:2025ype}, we estimate the SNR of GW250114 to be $\sim 77$. This makes it the loudest GW detected to date; it is $\sim 3$ times louder than GW150914~\cite{LIGOScientific:2016aoc}, which was the first and loudest BBH detected prior to O4. GW250114 thus provides an excellent opportunity to explore fundamental questions about BBH formation and allow for stringent tests of general relativity (GR).

In this work, we analyze a set of BBH simulations with properties similar to what we expect GW250114 to have, to highlight new insights that may be gained from this event. We investigate how well key physical properties of the system can be measured, such as its components' masses and spins, as well as any potential eccentricity. We also perform tests of GR on these simulated signals to determine what constraints may be inferred from this event. These tests include a quasi-normal mode analysis of the post-merger signal (``Black hole spectroscopy''), a time-domain inspiral-merger-ringdown (IMR) consistency test, and a consistency test of sub-dominant modes evaluated over the entire observable signal (``BBH spectroscopy''). To perform these tests, we select a non-spinning numerical relativity (NR) simulation similar to GW150914 as our reference signal, with masses and distance rescaled to match the SNR of GW250114. We also generate simulations using the \texttt{SEOBNRv5EHM}~\cite{Gamboa:2024hli} (to study eccentricity), and \texttt{NRSur7dq4}~\cite{Varma:2019csw} and \texttt{IMRPhenomXPHM}~\cite{Pratten_2021} (to study spin) waveform models.

All simulations are performed in a zero-noise realization of the Hanford and Livingston detectors, which were the observatories that detected GW250114~\cite{GraceDB}. We use the power spectral density (PSD) of each detector around the time of the event to evaluate the likelihood function while doing Bayesian inference. The PSDs are obtained by digitizing publicly available plots of the detectors' median PSD measured over the day of the event~\cite{DetectorStatus}. All analyses are conducted with the latest version of PyCBC Inference~\cite{Biwer:2018osg,alex_nitz_2024_10473621}.

\section{Source Characterization}
 
To establish a baseline for comparison, we use a simulation of a BBH merger from numerical relativity (NR) as our reference signal. Specifically, we use \texttt{SXS:BBH:3984} from the latest release of the Simulating eXtreme Spacetimes (SXS) catalog~\cite{scheel2025sxscollaborationscatalogbinary}. This simulation corresponds to a non-eccentric BBH merger with a mass ratio $q \equiv m_{1}/m_{2} = 1.5$ and with nearly zero spins on each component. It also exhibits high numerical accuracy and convergence. To check its numerical accuracy, we compute the mismatch between the highest and second-highest resolution used in the NR simulations. We find that the mismatch is $1.24 \times 10^{-5}$, indicating excellent numerical convergence consistent with previously reported results~\cite{scheel2025sxscollaborationscatalogbinary,Mitman_2025}. 

For our analysis, we rescale the total mass of the reference signal to $M=69.923\, M_{\odot}$, resulting in component masses of $m_1=41.954\, M_{\odot}$ and $m_2=27.969\, M_{\odot}$. We set the luminosity distance to $D_L=440\, \rm Mpc$ and the inclination angle $\iota = 36.6^\circ$, which corresponds to the most probable orientation. These masses, distance, and inclination are chosen to yield a network SNR of $\sim 77$~\cite{Akyuz:2025ype}. The sky location and luminosity distance are obtained from the maximum of the posterior distribution in the publicly available FITS files on the Gravitational-Wave Candidate Event Database (GraceDB)~\cite{GraceDB}, as produced by \texttt{Bilby}~\cite{Ashton:2018jfp}.

\begin{figure}
     \centering
     \includegraphics[width=\linewidth]{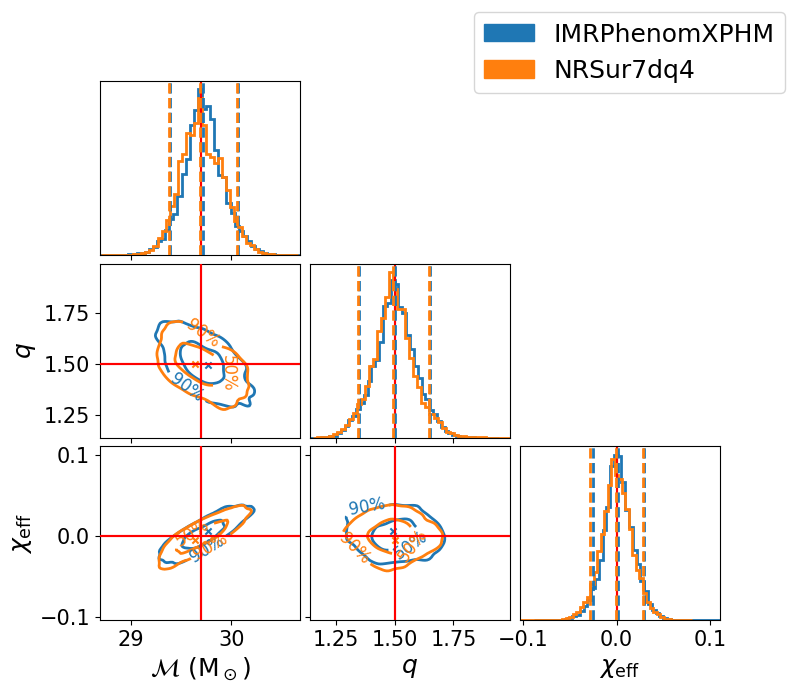}
     \caption{Marginal posterior distributions on chirp mass $\mathcal{M}$, mass ratio $q$, and effective spin $\chi_{\rm eff}$ of the reference NR simulation, obtained using \texttt{IMRPhenomXPHM}~\cite{Pratten_2021} (blue) and \texttt{NRSur7dq4}~\cite{Varma:2019csw} (orange). The $50\%$ and $90\%$ credible regions are shown in the 2D marginal posterior plots; the middle $90\%$ is indicated by the vertical dashed lines in the 1D marginal posteriors. Red lines indicate the injected values. Both models accurately recover the injected values.}
     \label{fig:nr_imr_nrsur_comp}
\end{figure}

To evaluate systematic uncertainties arising from waveform modeling, we perform Bayesian inference on this NR injection using the \texttt{IMRPhenomXPHM}~\cite{Pratten_2021} and \texttt{NRSur7dq4}~\cite{Varma:2019csw} waveform models. A comparison of the recovery of chirp mass $\mathcal{M}$, mass ratio $q$, and effective spin $\chi_{\rm eff}$ are shown in Fig.~\ref{fig:nr_imr_nrsur_comp} and Table~\ref{tab:recovered_parameters}. We find excellent agreement between the two models, and both models accurately recover the injected parameters, consistent with expectations for non-spinning signals~\cite{Pratten_2021,Varma:2019csw}. If GW250114 is non-spinning, systematic errors in waveform models should therefore be small relative to statistical errors.

\begin{table*}
\centering
\begin{tabular}{lccccccc}
\hline
\addlinespace
Simulation  & Model & \texttt{$\mathcal{M}$ [$M_\odot$]} & \texttt{$M_{\mathrm{tot}}$ [$M_\odot$]} &  \texttt{$\chi_{\mathrm{eff}}$} & \texttt{$\chi_p$} \\
\addlinespace
\hline
\addlinespace
\addlinespace
Reference & \texttt{SXS:BBH:3984} & 29.699 & 69.923 & 0 & 0 \\
High Precession & \texttt{NRSur7dq4} & 29.699 &  69.923 & $-3.012 \times10^{-6}$ & 0.9 \\
Moderate Precession & \texttt{NRSur7dq4} & 29.699 & 69.923 & 0.262 & 0.636 \\
Anti Aligned Spin & \texttt{NRSur7dq4} & 37.165 & 87.5 & -0.680 & $2.87 \times10^{-9}$ \\
Aligned Spin & \texttt{NRSur7dq4} & 23.361 & 55 & 0.680 & 0 \\
Non-Spinning & \texttt{NRSur7dq4} & 29.699 & 69.923 & 0 & 0 \\
\addlinespace
\hline
\end{tabular}
\caption{Injected chirp mass $\mathcal{M}$, total mass $M_{\mathrm{tot}}$, effective spin $\chi_{\rm eff}$, and $\chi_{p}$ of the simulations. All simulations have a mass ratio of 1.5. Values reported for chirp mass $\mathcal{M}$ and total mass $M_{\mathrm{tot}}$ are in the detector frame. The mass of the aligned and anti-aligned spin simulations differ from the other simulations in order to keep the SNR $\approx 77$, which is the expected SNR of GW250114~\cite{Akyuz:2025ype}.}
\label{tab:injection_values}
\end{table*}

\begin{table*}
\centering
\begin{tabular}{lcccccccc}
\hline
\addlinespace
Simulation  & Model & \texttt{$\mathcal{M}$ [$M_\odot$]} & \texttt{$q$} & \texttt{$M_{\mathrm{tot}}$ [$M_\odot$]} &  \texttt{$\chi_{\mathrm{eff}}$} & \texttt{$\chi_p$} \\
\addlinespace
\hline
\addlinespace
\addlinespace
Reference & \texttt{IMRPhenomXPHM} & $29.72^{+0.35}_{-0.32}$ & $1.50^{+0.15}_{-0.15}$ & $69.98^{+0.84}_{-0.76}$ & $0.001^{+0.028}_{-0.027}$ & $0.070^{+0.146}_{-0.056}$ \\
Reference & \texttt{NRSur7dq4} & $29.69^{+0.37}_{-0.31}$ & $1.49^{+0.16}_{-0.15}$ & $69.91^{+0.84}_{-0.81}$ & $0.000^{+0.028}_{-0.028}$ & $0.062^{+0.121}_{-0.049}$ \\
High Precession & \texttt{NRSur7dq4} & $29.85^{+0.41}_{-0.32}$ & $1.48^{+0.16}_{-0.15}$ & $70.27^{+0.84}_{-0.82}$ &  $0.009^{+0.039}_{-0.038}$ & $0.85^{+0.10}_{-0.11}$ \\
Moderate Precession & \texttt{NRSur7dq4}  & $29.67^{+0.35}_{-0.24}$ & $1.46^{+0.14}_{-0.15}$ & $69.67^{+1.16}_{-0.91}$ & $0.263^{+0.037}_{-0.025}$ & $0.59^{+0.25}_{-0.19}$ \\
Anti Aligned Spin & \texttt{NRSur7dq4}  & $37.4^{+1.0}_{-1.0}$ & $1.49^{+0.18}_{-0.14}$ & $88.00^{+1.9}_{-1.6}$ & $-0.669^{+0.065}_{-0.067}$ & $0.20^{+0.17}_{-0.12}$ \\
Aligned Spin & \texttt{NRSur7dq4}  & $23.31^{+0.18}_{-0.16}$ & $1.60^{+0.26}_{-0.23}$ & $55.38^{+1.09}_{-0.89}$ & $0.658^{+0.041}_{-0.041}$ & $0.129^{+0.144}_{-0.083}$ \\
Non-Spinning & \texttt{NRSur7dq4} & $29.73^{+0.29}_{-0.23}$ & $1.49^{+0.13}_{-0.14}$ & $69.97^{+0.76}_{-0.71}$ & $0.002^{+0.025}_{-0.024}$ & $0.060^{+0.120}_{-0.049}$ \\
\addlinespace
\hline
\end{tabular}
\caption{Recovered parameters with 90\% credible intervals for the simulations shown in Table~\ref{tab:injection_values}.  Here, the Model column refers to the waveform model used to recover the simulation. Values reported for chirp mass $\mathcal{M}$ and total mass $M_{\mathrm{tot}}$ are in the detector frame.}
\label{tab:recovered_parameters}
\end{table*}

\subsection{Eccentricity}
Orbital eccentricity describes the deviation of a binary's orbit from a perfect circle, and produces a gravitational wave with distinctive phase and amplitude modulations~\cite{PhysRev.136.B1224}. Systems with non-negligible eccentricity can enter the detector band with residual eccentricity, which may be measurable and provide insights for the formation mechanisms of such binaries. In particular, high eccentricity is often associated with dynamical formation channels, such as interactions in dense stellar environments~\cite{Zevin_2021,Romero-Shaw:2020thy,DallAmico:2023neb,Gupte:2024jfe}. 

To investigate the measurability of eccentricity for an event like GW250114, we perform an injection study using the new eccentric aligned spin waveform model \texttt{SEOBNRv5EHM}~\cite{Gamboa:2024hli}. We inject three dominant mode signals using the same intrinsic and extrinsic parameters as the reference injection, but with three different eccentricities $e =[0, 0.05, 0.1]$ at a reference frequency of 20 Hz, and a fixed radial anomaly of $l =\pi$, which is the angle that describes the position of the object in the eccentric orbit. Since the model does not account for precession, we fix the aligned spins to be 0. For Bayesian analysis, we use a uniform prior distribution for eccentricity, a uniform distribution for aligned spins, and a uniform angle on radial anomaly. Both injections and subsequent Bayesian analyzes are performed using \texttt{SEOBNRv5EHM}, allowing us to assess how well eccentricity can be constrained in each case.

\begin{figure}
    \centering
    \includegraphics[width=0.85\linewidth]{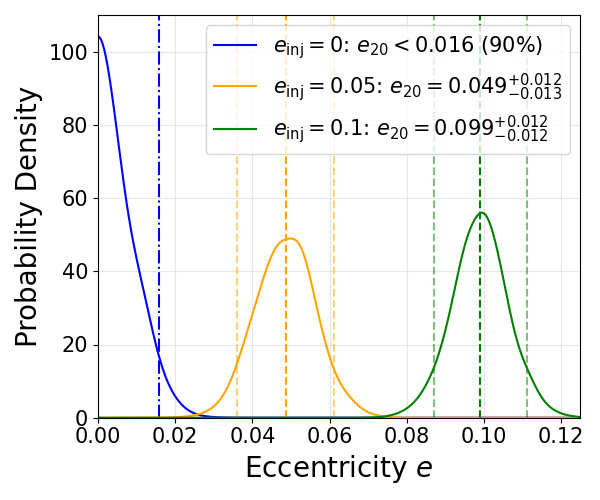}
    \caption{Posterior distributions of orbital eccentricity for three injection studies of GW250114. The blue curve corresponds to the $e = 0$ injection, the orange to $e = 0.05$, and the green to $e = 0.1$ for a reference frequency of 20Hz. Each posterior is recovered using the \texttt{SEOBNRv5EHM} waveform model. Eccentricity cannot be bounded away from zero in the $e = 0$ case, while it is potentially measurable for the $e = 0.05$ and $e = 0.1$ injections.}
    \label{fig:ecc}
\end{figure}

We present our results in Fig.~\ref{fig:ecc}. We find that eccentricity is not measurably different from zero in the $e = 0$ injection, with the 90\% credible upper bound constrained to $e < 0.016$. For the $e = 0.05$ injection, the recovered posterior yields $e = 0.049^{+0.012}_{-0.013}$, tightly centered around the injected value and clearly excluding zero. This indicates that eccentricity at the level of 0.05 is distinguishable from a quasi-circular signal with current methods. The $e = 0.1$ injection is recovered as $e = 0.099^{+0.012}_{-0.012}$, also excluding zero and showing a strong deviation from circularity. These results suggest that eccentricities of $e \gtrsim 0.05$ may be detectable, while lower values, such as those consistent with zero, are not well measured.

\subsection{Spin}

\begin{figure*}[ht]
    \centering
    \includegraphics[width=\linewidth]{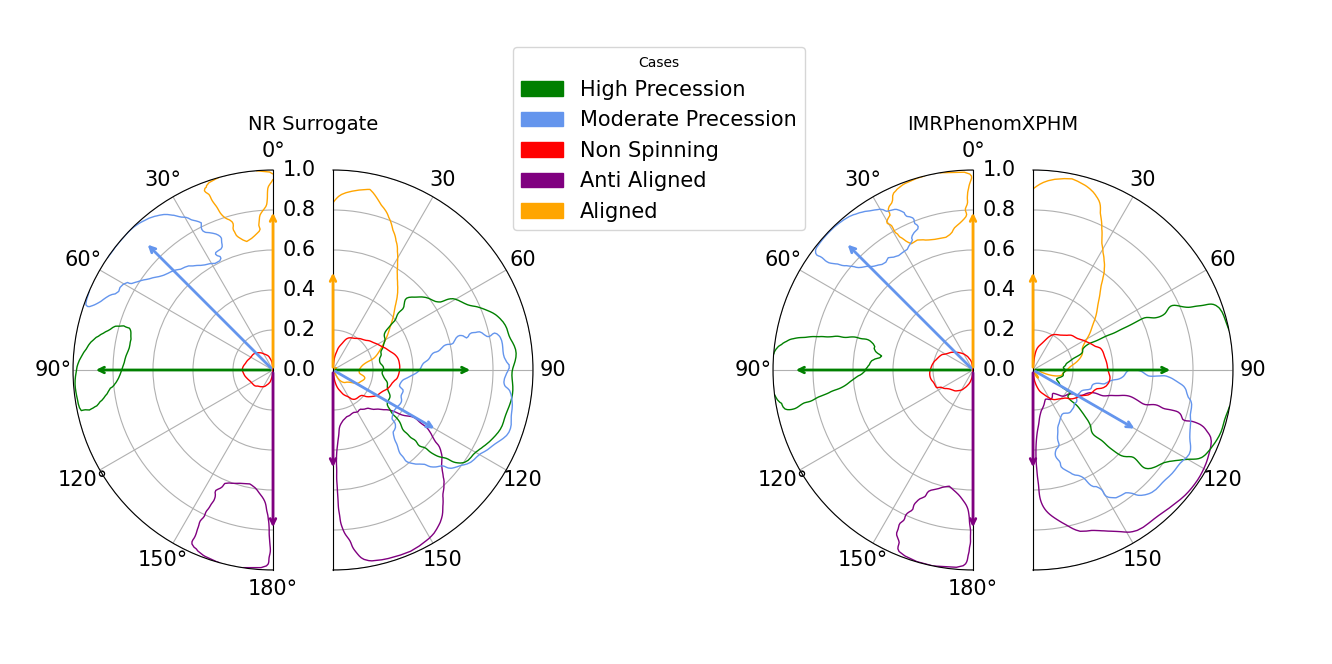}
    \caption{90\% credible regions of the posterior distributions for the spin magnitude and polar angle. Each polar plot is split into two hemispheres: the left hemisphere represents the primary spin, and the right hemisphere represents the secondary spin. The left panel shows results from the NR Surrogate analysis using a common injection model, while the right panel shows results from the \texttt{IMRPhenomXPHM} analysis using the same injection model. Colored contours correspond to different spin configurations as indicated in the legend, and arrows indicate the amplitude and orientation of the injected values.}
    \label{fig:spin}
\end{figure*}

The formation history of the BBH affects the orientation and magnitude of the spin of the initial black holes. For black holes formed via stellar collapse, the spin magnitude is influenced by the core-envelope coupling~\cite{Fuller:2019sxi}. In contrast, black holes resulting from previous mergers are expected to have dimensionless spin magnitudes around 0.7~\cite{Gerosa:2017kvu}. The spin alignment depends on the formation environment of the binary, due to conservation of angular momentum. For instance, in dynamical environments, interactions with surrounding objects lead to an isotropic distribution of spin orientations, and spin precession is therefore expected \cite{Phukon:2025yva}. On the other hand, in isolated environments, the spins are expected to align with the orbital angular momentum, and precession is typically not observed~\cite{Phukon:2025yva}.

To establish what spin constraints may be expected for GW250114 (and thus what can be gleaned about its formation history), we simulate five different scenarios: (1) a non-spinning system; (2) a system with aligned spin; (3) a system with anti-aligned spin; (4) a system with high precession; and (5) a system with moderate precession. The corresponding values of effective spin ($\chi_{\mathrm{eff}}$) and processing spin ($\chi_p$) values for each scenario are listed in Table~\ref{tab:recovered_parameters}. 

The results are shown in Fig.~\ref{fig:spin}. In the aligned spin case, the primary spin amplitude is constrained to be greater than $\sim 0.65$ for both waveform models. In the high-precession case, the primary spin amplitude is bounded below by $\sim 0.75$ for the \texttt{NRSur7dq4} and $\sim 0.5$ for the \texttt{IMRPhenomXPHM} analyses. For moderate precession, we can bound the primary spin amplitude to be higher than $\sim 0.6$ for \texttt{NRSur7dq4} and $\sim 0.7$ for \texttt{IMRPhenomXPHM}. For the anti-aligned case, we can bound the primary spin amplitude to be higher than $\sim 0.6$ for both waveform models. For these analyses, secondary spin bounds are not significant. 

For the non-spinning case with \texttt{NRSur7dq4} injection, the primary spin is bounded lower than $\sim 0.15$ and $\sim 0.3$ for the secondary spin while for \texttt{IMRPhenomXPHM} injection the primary spin is bounded lower than $\sim 0.2$ and the secondary spin is bounded lower than $\sim 0.4$. Even if GW250114 only has moderate precession, it should be possible to bound the magnitude of the secondary object's spin away from zero.

\section{Tests of General Relativity}

\subsection{Quasinormal mode detectability}
\label{sec:bhspec}
Gravitational-wave observations allow for high-precision tests of GR in the strong-field regime of gravity. For example, we can perform ``black hole spectroscopy" on the post-merger ``ringdown" of binary black hole (BBH) coalescence. The ringdown of a BBH merger GW can be expressed as a superposition of quasinormal modes (QNM) specified by two angular indices $\ell \geq 2$, $m \leq |\ell|$, and an overtone number $n \geq 0$~\cite{Vishveshwara:1970zz,Chandrasekhar:1975zza}. According to the no-hair theorem, the frequency and damping time measured for each mode should be maximally determined by the mass and spin of the remnant BH. Inconsistencies between the measured QNMs and final mass and spin would imply inconsistencies with Kerr spacetime as predicted by GR~\cite{Dreyer:2003bv}.

We can test the no-hair theorem directly if a GW post-merger signal contains at least two detectable QNMs. With the exception of some highly precessing systems~\cite{Zhu:2023fnf}, the $(\ell, m, n) = (2,2,0)$ QNM is expected to be the dominant ringdown mode emitted by the black hole formed by a BBH merger. Several studies have claimed detection of subdominant modes in LVK data. Some authors claimed detection of a $(2,2,1)$ overtone in the ringdown of GW150914 and subsequently found agreement with the no-hair theorem~\cite{Isi:2019aib,Finch:2022ynt,Wang:2023xsy}. Other analyses probed for the detectability of subdominant fundamental modes ($\ell, m \neq 2, n=0$) in the ringdown of GW190521~\cite{Capano:2021etf,Siegel:2023lxl}.

Each analysis that claims subdominant mode detection --- and even when the QNM model becomes a valid description of the post-merger --- is hotly debated in the literature~\cite{Cotesta:2022pci,Bhagwat:2019dtm,Jaramillo:2020tuu,Ma:2022wpv,Cheung:2022rbm,Mitman:2022qdl,Nee:2023osy,Baibhav:2023clw,London:2014cma}. At least some of this contention is due to the choices of coalescence time ($t_c$) and sky location, which have traditionally been fixed due to technical challenges~\cite{Correia:2023ipz,Finch:2022ynt,Wang:2023xsy}.

Correia \& Capano~\cite{Correia:2023ipz} developed methods to address these challenges and marginalize over the sky location and $t_c$ uncertainty while performing a QNM analysis of the post-merger signal. The authors applied these methods to an analysis of GW150914 and found that the data was uninformative as to the observability of the $(2,2,1)$ mode, with a Bayes factor in favor of the overtone model $Z_{220+221)}/Z_{220}$ of only 1.15~\cite{Correia:2023bfn}. They concluded that the SNR of GW150914 was too low: the resultant timing uncertainty was too large to confidently detect the overtone.

The large SNR of GW250114 makes overtone detection more likely. To probe what may be extracted from that event, we apply the same methods of sky and time marginalization to our non-spinning reference simulation (\texttt{SXS:BBH:3984}). A full description of our methods is outlined in~\cite{Correia:2023ipz}. We test three QNM models: a baseline $(2,2,0)$ model, a $(2,2,0)+(2,2,1)$ model, and a $(2,2,0)+(2,2,1)+(2,2,2)$ model. 

\begin{figure}
    \centering
    \includegraphics[width=1.02\linewidth]{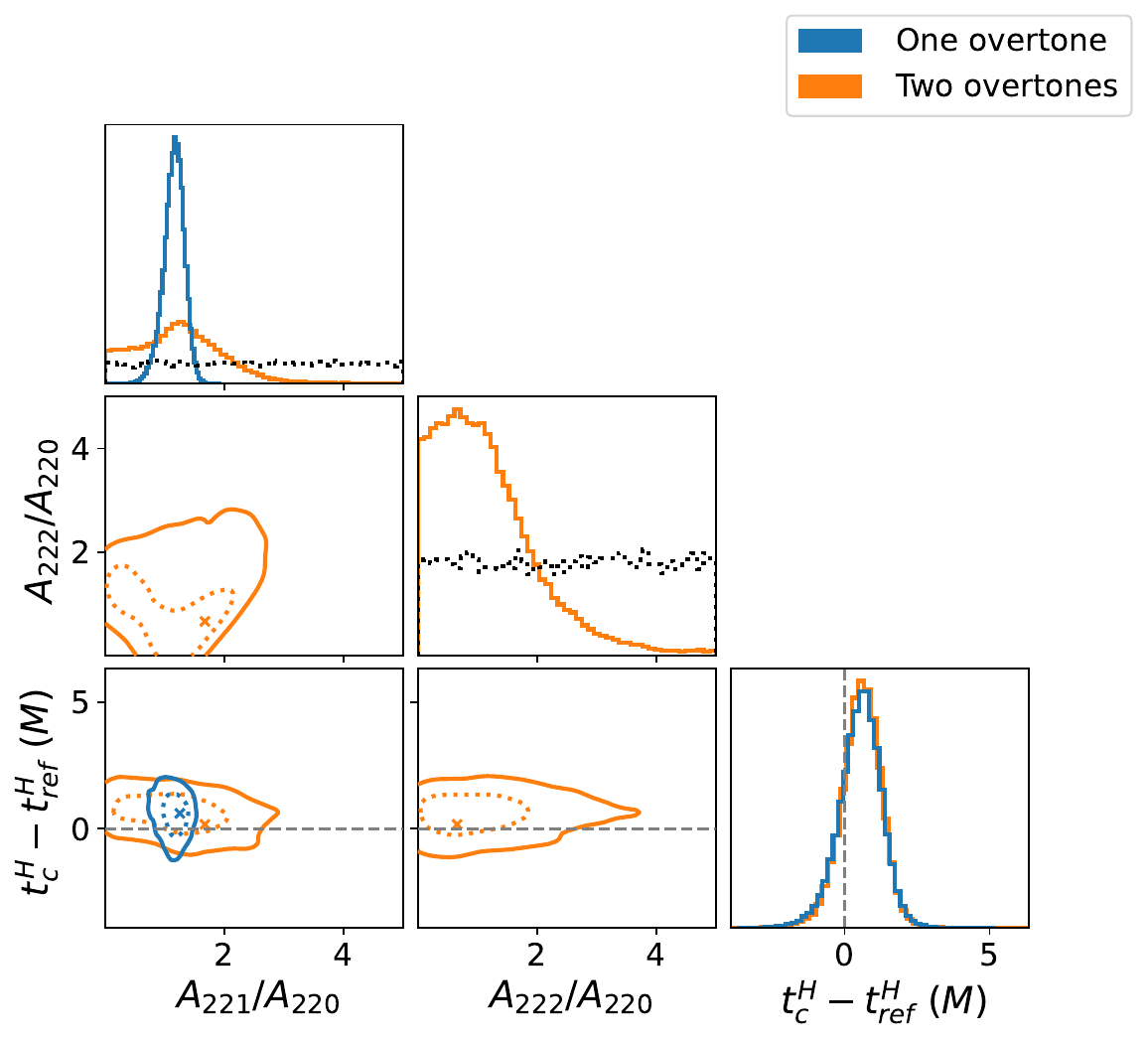}
    \caption{The relative overtone amplitude and Hanford $t_c$ posteriors for our QNM analyses. Results from the one-overtone ($(2,2,0) + (2,2,1)$ ringdown) and two-overtone ($(2,2,0) + (2,2,1) + (2,2,2)$ ringdown) models are indicated with blue and orange lines, respectively. Dotted and solid contours indicate the 50th and 90th percentiles of the posteriors, respectively. The maximum likelihood point in each 2D posterior is indicated with an X. Values of $t_c$ are expressed relative to the injected Hanford $t_c$ value in units of the final mass of our reference injection. A dashed gray line indicates the reference $t_c$ of our injection. A dotted black line on each 1D histogram indicates the prior for the corresponding parameter.}
    \label{fig:overtone}
\end{figure}

We calculate the Bayes factor $\mathcal{B}_{QNM} = Z_{QNM}/Z_{220}$ using the evidence $Z$ for each model. For the one-overtone model, we obtain $\mathcal{B}_{221} = 2492$. Similarly, for the two-overtone model, we obtain $\mathcal{B}_{221+222} = 724$. These values indicate ``decisive''~\cite{Kass:1995loi} support for our injection to contain at least one detectable overtone. Figure~\ref{fig:overtone} shows the amplitude and coalescence time posteriors for our analyses. Our one-overtone model posterior exhibits considerable support for $A_{221}$ away from zero, thereby agreeing with the Bayes factor interpretation. Meanwhile, our two-overtone model exhibits more support for $A_{222}$ away from zero and $A_{221}$ at zero, and neither amplitude has as much support away from zero as the one-overtone model. These results suggest that one overtone may be definitively detected in GW250114.

We do not expect sub-dominant fundamental modes ($\ell, m \neq 2; n=0$) to be detectable in GW250114. Fundamental modes are only expected to be observable in signals with high $q$ or precession~\cite{Zhu:2023fnf,Nobili:2025ydt,Siegel:2023lxl}. For our NR reference injection (with nonspinning progenitor BHs and $q \approx 1.5$), we estimate a total ringdown SNR of roughly 18, while the SNR of all other modes is around 2.1. Furthermore, none of the fundamental modes exceed an SNR of $\sim1.8$. This indicates a low probability of detecting fundamental modes in our reference signal. However, if the real data for GW250114 exhibits evidence for a high mass ratio ($q \gtrsim 4$) or precession effects, a fundamental QNM analysis will be warranted.

\subsection{Time-domain IMR consistency}
By independently measuring binary parameters from the pre- and post-merger signal, it is possible to test the consistency of observed events with the laws of black hole thermodynamics~\cite{Bardeen:1973gs}. Notably, we can test Hawking's area theorem~\cite{Hawking:1971tu}, which states that after the coalescence of two black holes, the area of the final event horizon, $A_f$, must be greater than or equal to the sum of the horizon areas of the progenitor black holes, $A_1$ and $A_2$: 
\begin{equation}
    A_1 + A_2 \equiv A_i \leq A_f\,.
    \label{eq:1}
\end{equation}
It is a direct consequence of the second law of thermodynamics that the entropy of a black hole is proportional to its area~\cite{Bardeen:1973gs}. If $M$ is the mass and $\chi = J/M^2$ is the dimensionless spin of the black hole, then the area of a black hole is given by
\begin{equation}
    A = 8\pi M^2 (1 + \sqrt{1 - \chi^2})\,.
    \label{eq:2}
\end{equation}

To test the area theorem, we use the ratio between the measured and expected changes in areas~\cite{Kastha:2021chr}
    \begin{align}
    \textit{R} = \frac{\Delta A_{\text{measured}}}{\Delta A_{\text{expected}}} = \frac{A_{f,\text{measured}} - A_i}{A_{f,\text{expected}} - A_i}\,.
    \label{eq:3}
\end{align}

$A_{f,measured}$ is measured from the final mass and spin estimated in the post-merger analysis. The component masses and spins prior to the merger are estimated from the inspiral (pre-merger) phase and then used to predict the final mass and spin via a NR fitting model. These predicted values are used to calculate the expected final horizon area $A_{f,\text{expected}}$ using Eq.~\eqref{eq:2}. As $A_{f,\text{expected}}$ is derived from the NR model using the initial horizon areas $A_1$ and $A_2$, the denominator of the test statistic $\textit{R}$ is positive definite. Therefore, if $\textit{R} < 0$, it means that $A_{f,\text{measured}} < A_i$ and suggests a violation of the black hole area theorem. Furthermore, $\textit{R} = 1$ indicates that the signal is consistent with GR. 

\begin{figure}
    \centering
    \includegraphics[width=1.02\linewidth]{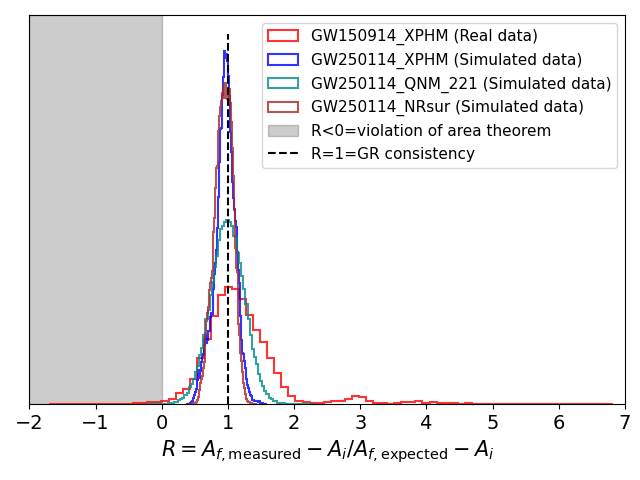}
    \caption{Comparison area ratio plot $\textit{R}$ between GW$150914$ and GW$250114$. GW250114 with \texttt{IMRPhenomXPHM} and \texttt{NRSur7dq4} approximants (blue and brown distributions respectively) show that a large fraction of posterior samples $(P(R>0) = 100.00\%)$ lie in the region $\textit{R} > 0$, indicating strong consistency with the area theorem in comparison with red distribution (GW$150914, P(R>0) = 99.36\%)$~\cite{Correia:2023ipz}. The dark cyan distribution shows that including the 221 QNM mode as a ringdown approximant leads to a tighter distribution, consistent with expectations from QNM-based analyses.}
    \label{fig:area_theorem}
\end{figure}  

$\Delta A_{\text{measured}}$ and $\Delta A_{\text{expected}}$ in Eq.~\eqref{eq:3} depend on the initial and final values of masses and spins, determined from the parameter estimation of the pre and post merger part of the signal, respectively. Due to computational challenges in likelihood evaluation, earlier studies estimated these parameters while fixing sky location and $t_c$~\cite{Kastha:2021chr,Isi:2020tac}. However, this can lead to biases in the values of the measured parameters~\cite{Cabero:2017avf} and underestimate the statistical uncertainties~\cite{Correia:2023ipz}. The method for marginalizing over sky location and $t_c$ used in Correia et al.~\cite{Correia:2023bfn} for black hole spectroscopy can also be used for testing the area theroem~\cite{Correia:2023ipz}. This is because marginalizing over these parameters involves simultaneously yet independently modeling the pre- and post-merger of the signal, allowing only sky-location and $t_c$ to be shared between them. The initial areas $A_1$ and $A_2$ can therefore be measured from the pre-merger signal, while $A_f$ is (independently) measured from the post-merger. We consider the same approach of marginalization of the sky location and $t_c$ to recover parameters and verify the area theorem.

The results of the area theorem test applied to the reference NR signal are shown in Fig.~\ref{fig:area_theorem}. We perform the test using \texttt{NRSur7dq4} as the model for both the pre- and post-merger signal, and separately a test using \texttt{IMRPhenomXPHM} for both the pre- and post-merger signal. We also plot the result using the $(2,2,0)+(2,2,1)$ QNM model for the post-merger (with \texttt{NRSur7dq4} used for the pre-merger), which was our most favored model from the BH spectroscopy test (Sec.~\ref{sec:bhspec}). The \texttt{IMRPhenomXPHM} and \texttt{NRSur7dq4} models yield consistent results, with $R = 0.96^{+0.23}_{-0.30}$ and $R = 0.94^{+0.20}_{-0.25}$, respectively. As can be seen in the plot, the QNM result is less constrained, but still centered on the expected value of $R=1$. This is consistent with what was observed in GW150914~\cite{Correia:2023ipz}. The potential constraints from GW250114 are significantly tighter than GW150914, however: that event yielded $R = 1.11^{+0.95}_{-0.63}$ using the \texttt{IMRPhenomXPHM} waveform model for both the pre- and post-merger phases~\cite{Correia:2023ipz}. In particular, we find that the entirety of the posterior $R$ for GW250114 is greater than zero. This observation indicates that it should be possible to confirm the area theorem (under the assumptions of this analysis) to very high confidence with GW250114. For comparison, $P(R>0)=99.36$ for GW150914 using \texttt{IMRPhenomXPHM} for the post-merger model~\cite{Correia:2023ipz}. 

\subsection{BBH spectroscopy}

GW signals can be decomposed using spin-weighted spherical harmonics as follows,
\begin{equation}
    h = h_{+} - ih_{\times} = \frac{1}{D_{L}}\sum_{(l,m)}{}^{-2}Y_{lm}(\theta) e^{i(\psi_{lm}(\theta) + \Phi_{c})}\,.
\end{equation}

Each of these modes should be dependent only on the intrinsic parameters of the binary - the two masses and spins. Thus, measurements of 2 or more harmonics would provide a way to measure any non-GR degrees of freedom. One way of doing this is using ``BBH Spectroscopy"~\cite{Capano:2020dix}. In this test, we add non-GR parameters for the higher modes and treat them as free parameters to estimate. We look at deviations of the chirp mass, symmetric mass ratio, and coalescence phase from the dominant (2,2) mode as our non-GR degrees of freedom.

\begin{align}
    \mathcal{M}_{lm} &= \mathcal{M}_{22}(1 + \delta\mathcal{M}_{lm})\,, \\
    \eta_{lm} &= \eta_{22}(1 + \delta \eta_{lm})\,, \\
    \Phi_{c,lm} &= \Phi_{c,22} + \Delta \Phi_{c,lm}\,.
\end{align}

For our reference injection, the dominant modes in terms of SNR are the (2,2) mode with SNR = 86.92 followed by the (3,3) modes with SNR = 6.51. There also exists a small contribution from the (2,1) mode with an SNR = 2.36. We perform BBH spectroscopy using only a higher (3,3) mode. We use \texttt{IMRPhenomXPHM} as the template waveform. We fix the spins to the reference injection values, as well as all extrinsic parameters except for inclination, since they affect each mode in the same way. We recover the masses, coalescence phase, and the deviations in them for the (3,3) mode.

\begin{figure}
    \centering
    \includegraphics[width=\linewidth]{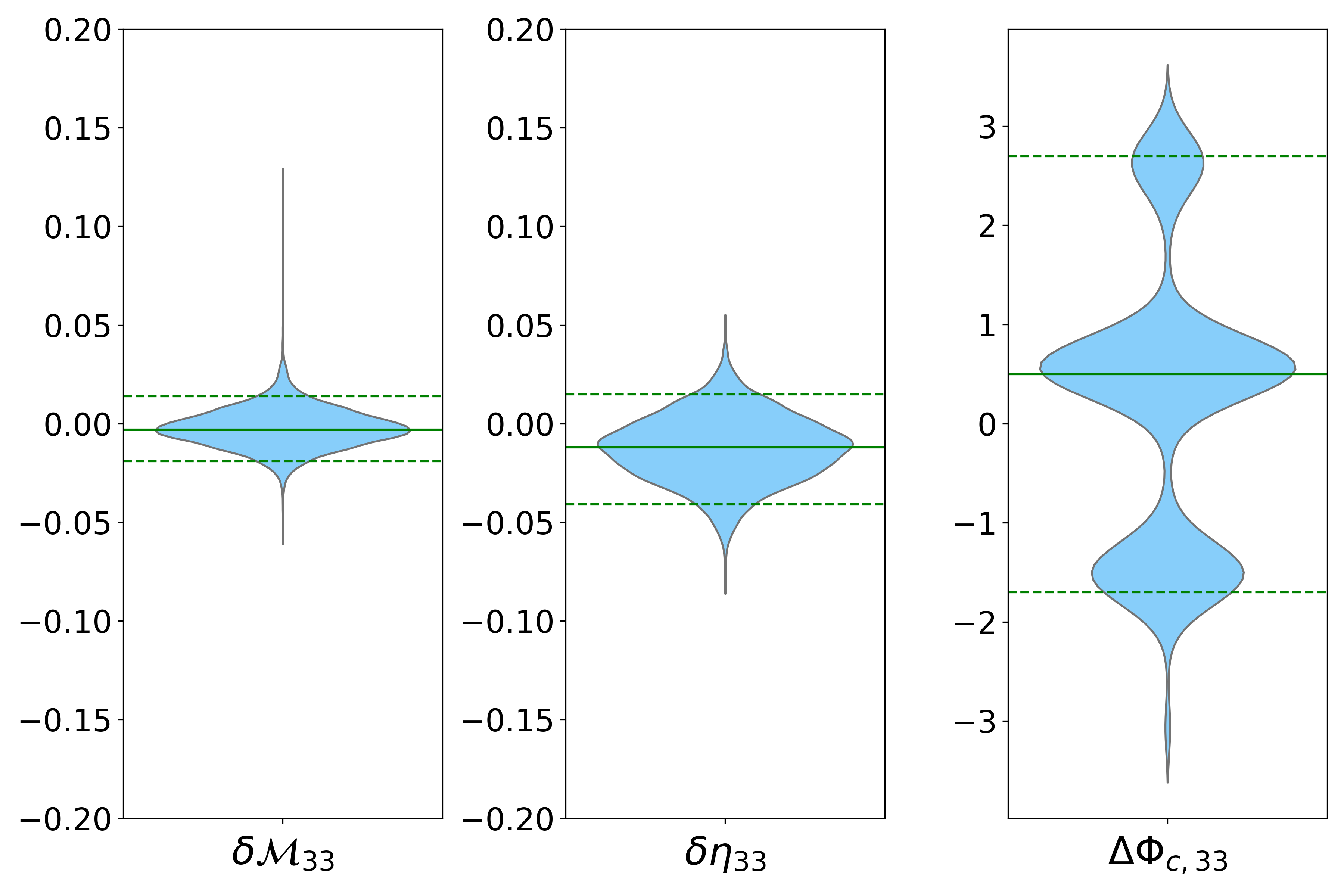}
    \caption{Marginalized posteriors for the fractional deviations in chirp mass ($\delta\mathcal{M}_{33}$) symmetric mass ratio ($\delta \eta_{33}$), and the absolute deviation in the phase ($\Delta \Phi_{c,33}$) of the (3,3) mode from the (2,2) mode, with the median values (solid green) and credible intervals (dotted green). The results indicate consistency with GR within a $90 \%$ credible interval.}
    \label{fig:spectroscopy}
\end{figure}

The distributions for fractional deviations in the (3,3) mode are shown in Fig.~\ref{fig:spectroscopy}. We see that the deviations in the mass parameters are $\delta\mathcal{M}_{33} = -0.005^{+0.017}_{-0.016}$ and $\delta \eta_{33} = -0.012^{+0.027}_{-0.029}$, are consistent with the predictions of GR. The measurement in the deviation in the coalescence phase does not provide a check of GR, but is consistent with (3,3) mode $\sim e^{i3\Phi_c}$, as the $90\%$ credible interval, correspond to angles separated by $ \sim 120^{0}$

\section{Conclusions}

Dynamical formation of BBHs is hypothesized to occur in dense environments. BBHs formed in this manner can potentially have non-zero eccentricity~\cite{Zevin:2021rtf} when entering the GW detector band, non-aligned spins, and relatively high spin values~\cite{Rodriguez:2019huv,Fuller:2019sxi,Fishbach:2017dwv}. Our study shows that it may be possible to identify these effects in GW250114. Regarding eccentricity, we found that it should be possible to measure a non-zero eccentricity at $\refFreq\,$Hz in GW250114 if it has $e_{\refFreq} \gtrsim 0.05$. The dynamical formation channel also implies the isotropic spin orientation. We find that even in the case of moderate precession (i.e., with $\chi_p = 0.636$), it should be possible to exclude a non-spinning or aligned-spin scenario at the SNR of GW250114. 

One of the most notable results of our study is that it should be possible to decisively measure an overtone of the dominant $(2,2,0)$ quasi-normal mode in the post-merger signal of GW250114, even when marginalizing over timing uncertainty. Previous detection claims of the overtone in GW150914 have been controversial due to the relatively large timing uncertainty~\cite{Isi:2019aib,Cotesta:2022pci,Wang:2023xsy,Correia:2023bfn}. The SNR of GW250114 means that an overtone detection should be more decisive, at least if it is non-spinning. Unfortunately, despite its overall large SNR, we do not believe that any sub-dominant fundamental QNM modes will be detectable with this event.

Even if GW250114 has a mass ratio as low as $\sim 1.5$, it should be possible to measure and place constraints on deviations of the $(3,3)$ mode from GR, as shown in Fig.~\ref{fig:spectroscopy}. We also find that it should be possible to obtain excellent constraints on Hawking's area theorem. We note, however, that the area-theorem constraint presented here is obtained with no gap in time between the pre-merger and post-merger measurement. A more rigorous test would be to measure the change in area between the early inspiral and the post-merger. We will pursue this in a future study.

Our analysis suggests that GW250114 has the potential to offer many new insights into the formation of BBHs and into the fundamental physics that govern them. We look forward to the public data release of GW250114 so that we can confirm our predictions.

\section{Data availability}

Data supporting this work, including HDF files containing posterior samples, configuration files needed to replicate results, and digitized PSD files, are openly available at~\cite{akyuz_2025_16738833}.

\acknowledgments 
AHN, KK, and KS acknowledge support from NSF grant PHY-2309240. CDC and AC acknowledge support from NSF grant PHY-2309356.
 
This research has used data or software obtained from the Gravitational Wave Open Science Center (gwosc.org), a service of the LIGO Scientific Collaboration, the Virgo Collaboration, and KAGRA. This material is based upon work supported by NSF's LIGO Laboratory which is a major facility fully funded by the National Science Foundation, as well as the Science and Technology Facilities Council (STFC) of the United Kingdom, the Max-Planck-Society (MPS), and the State of Niedersachsen/Germany for support of the construction of Advanced LIGO and construction and operation of the GEO600 detector. Additional support for Advanced LIGO was provided by the Australian Research Council. Virgo is funded, through the European Gravitational Observatory (EGO), by the French Centre National de Recherche Scientifique (CNRS), the Italian Istituto Nazionale di Fisica Nucleare (INFN) and the Dutch Nikhef, with contributions by institutions from Belgium, Germany, Greece, Hungary, Ireland, Japan, Monaco, Poland, Portugal, Spain. KAGRA is supported by Ministry of Education, Culture, Sports, Science and Technology (MEXT), Japan Society for the Promotion of Science (JSPS) in Japan; National Research Foundation (NRF) and Ministry of Science and ICT (MSIT) in Korea; Academia Sinica (AS) and National Science and Technology Council (NSTC) in Taiwan.

\clearpage
\bibliography{references}
\end{document}